\documentclass[final,3p,times,twocolumn]{elsarticle}

\usepackage{amssymb}
\usepackage{lineno,hyperref}
\hypersetup{hidelinks}
\usepackage{url}
\usepackage{amsmath}
\usepackage{multirow}
\usepackage{bm}
\usepackage[figuresright]{rotating}
\usepackage{bicaption}
\usepackage{booktabs,graphicx}
\usepackage{subfig}
\usepackage{algorithm}
\usepackage[]{algpseudocode}
\usepackage{amsfonts}
\usepackage{array}
\usepackage{booktabs}
\usepackage{multirow}
\usepackage{ragged2e}

\usepackage{array}
\usepackage{xspace}
\usepackage{adjustbox}
\usepackage{placeins}
\usepackage{wrapfig} 
\usepackage{pdflscape}
\usepackage{xcolor}
\usepackage{tikz}
\DeclareRobustCommand{\dnum}[1]{\tikz[baseline=(dnum-c.base)]{\node[circle,fill=black,text=white,inner sep=0pt,minimum size=10pt,font=\sffamily\bfseries\scriptsize](dnum-c){#1};}}

\DeclareRobustCommand{\litcite}[1]{{\color{black}\cite{#1}}}
\newlength{\litwidth}
\floatname{algorithm}{Procedure}

\usepackage{longtable}

\usepackage{CJKutf8}
\usepackage{orcidlink}
\usepackage{comment}

\modulolinenumbers[5]

\makeatletter
\let\ps@pprintTitle\ps@empty
\makeatother
\begin{document}

\begin{frontmatter}

\title{Integrating 3D Posture Neutralization with Statistical Shape Modeling for Head–Neck-Shoulder Ergonomic Design}

\author[sft]{Jinbo Luo\orcidlink{0009-0005-8793-4520}}
\author[hospit]{Hang Ren\orcidlink{0009-0001-7532-059X}}
\author[sft]{Yadie Yang\orcidlink{0000-0002-1985-7552}}
\author[mpu]{Jie Zhang\orcidlink{0000-0001-8219-5590}}
\corref{mycorrespondingauthor}
\ead{jpeter.zhang@mpu.edu.mo}

\address[sft]{School of Fashion and Textiles, The Hong Kong Polytechnic University, Hong Kong SAR, China}
\address[hospit]{ Shenzhen Pingle Orthopedic Hospital, Shenzhen, China}
\address[mpu]{Faculty of Applied Sciences, Macao Polytechnic University, Macau SAR, China}

\begin{abstract}
The design of ergonomic products and workspaces requires a deep understanding of human anatomy and posture. In particular, the head-neck-shoulder region is critical for many applications such as cervical orthoses, helmets, and other head-worn or neck-worn devices. This paper presents a framework that integrates posture neutralization with statistical shape modeling (SSM) for the head-neck-shoulder (HNS) region. We acquire a single three-dimensional (3D) scan of the head-neck-shoulder region from each of 200 Chinese adults in a natural posture for each subject. Every scan is then posture-neutralized before SSM. We fit a parametric head-neck model to each scan to separate cervical pose from anatomy, and reset the cervical pose to a common neutral configuration, so that the resulting model captures anatomical shape variation rather than posture-induced deformation. The resulting shape space yields a standardized template and population-wide modes of variation, providing baselines for parametric ergonomic design. We further discuss how the model supports population coverage analysis and the design of products that accommodate anatomical diversity while maintaining neutral cervical posture.
\end{abstract}


\end{frontmatter}

\begin{figure*}[t!]
\centering
\includegraphics[width=1\linewidth]{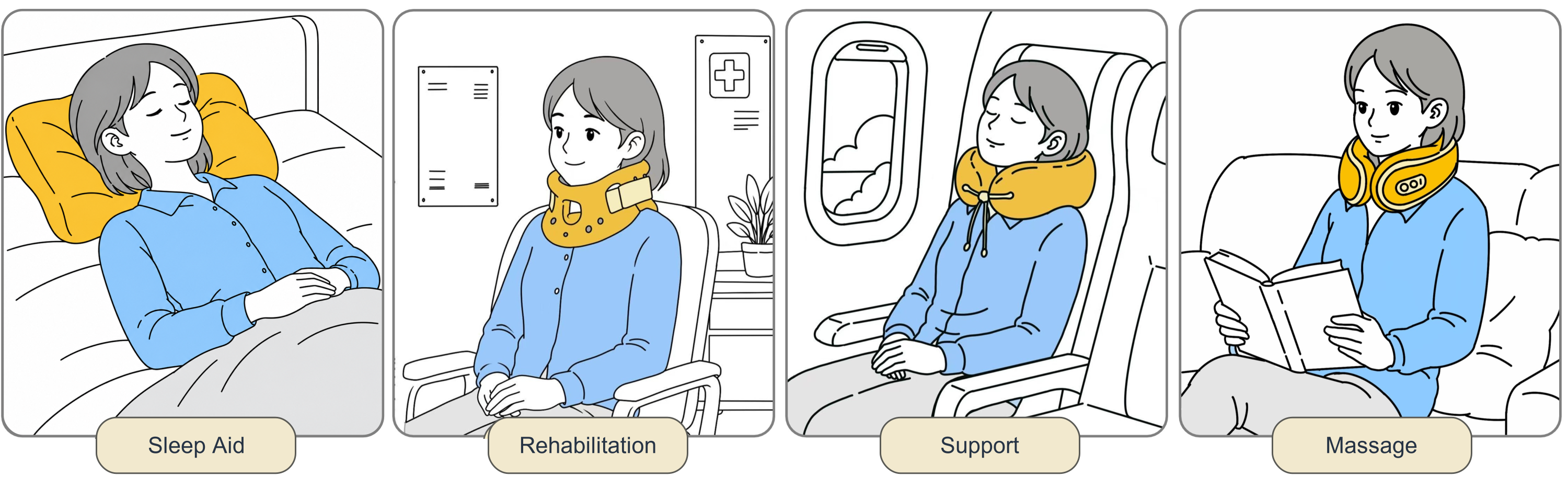}
\caption{Typical applications of 3D head-neck-shoulder morphological data. (This figure was created by using Procreate software.)}
\label{Figure_Application}
\end{figure*}

\section{Introduction}

Accurate acquisition and statistical modeling of 3D head-neck-shoulder morphological data is a prerequisite for the ergonomic design of cervical spine and related wearable products~\cite{zhang2023review,lacko2017product,kuo2020developing}. Without accurate morphological data, design decisions concerning fit, comfort, or safety often rely on empirical estimation, limiting the rigor of evidence-based customization~\cite{langley2018comfort,luximon2012sizing}.

In recent years, with the widespread adoption of information-based work and lifestyles, prolonged head-down use of electronic devices has led to the increasing prevalence of poor cervical postures such as forward head posture~\cite{fatharani2025prevalence,intaruk2025differences}. Studies have shown that when using small electronic devices such as smartphones, the neck flexion angle often reaches between $30^{\circ}$ and $45^{\circ}$~\cite{intaruk2025differences,lee2017effects}. Furthermore, simplified biomechanical estimates suggest that for every $15^{\circ}$ of forward head tilt, the equivalent load on the cervical spine may increase by approximately $12$ kg~\cite{hansraj2014assessment}, though this static model has been criticized for oversimplifying physiological loading conditions. Under such sustained high-load postures, the cervical musculoskeletal system is at elevated risk of chronic strain injury~\cite{mahmoud2019fhp,yang2023chronic}. The widespread prevalence of these poor postures has directly driven the rapid increase in cervical spine disorders. Research has shown that the incidence rate of cervical spondylosis in urban areas of China reached 405.65 per 100,000 person-years in 2017, with a significantly higher incidence in females (accounting for 60.58\%) and the highest incidence observed in the 50-59 age group~\cite{zhuo2026incidence}. Community-based surveys further estimate the symptomatic prevalence at around 13.76\% in Chinese adults~\cite{lv2018prevalence}. With the widespread adoption of information-based work habits and the aging population, cervical spondylosis exhibits a trend toward earlier onset and has become an increasingly serious public health concern. Against this backdrop, the demand for ergonomic pillows designed for daily cervical support has grown rapidly~\cite{lei2021pillow}, thereby making the statistical shape modeling (SSM)~\cite{cootes1995active} of 3D head-neck-shoulder shapes more significant and urgent. Beyond motivating the demand for cervical-supportive products, the wide range of cervical postures observed in daily life also creates a methodological challenge for SSM~\cite{salahzadeh2014assessment}. Scans acquired from subjects in different postures mix posture-induced deformation with true anatomical variation between individuals. Separating these two sources of variation is a prerequisite for any shape model intended to inform ergonomic design~\cite{danckaers2019posture}.  

Multiple standardized head-scanning databases have been built for different populations, such as CAESAR~\cite{harrison2002caesar}, NIOSH~\cite{zhuang2013head}, SizeChina~\cite{luximon2012chinese}, SizeChina-Hunan~\cite{wang2018digital}, Size-Korea~\cite{seo2020development}, Headspace~\cite{dai2020statistical}, LSFM~\cite{booth2018large}, and FaceScape~\cite{zhu2023facescape}. Established methods include template-based registration~\cite{allen2003space,amberg2007optimal} and statistical modeling of head shape using principal component analysis (PCA)~\cite{egger20203dmm,zhang2022age}. These methods support sizing-system development and parametric design of head-support products. Table~\ref{tab:hns_literature_review} summarizes regional studies by anatomical coverage, shape representation and posture treatment. Neck and neck-shoulder representations include measurement-based classifications, NURBS surfaces and mean meshes~\cite{wang2017classification,zhang2022neckshoulder,wang2021neckshoulder,baiamonte2024neck}.

\begin{table*}[t]
\centering
\color{black}
\captionsetup{font={small,color=black},labelfont={color=black},textfont={color=black}}
\caption{Comparison of head, neck and shoulder shape studies.}
\label{tab:hns_literature_review}
\fontsize{7.2}{8.6}\selectfont
\setlength{\tabcolsep}{1pt}
\renewcommand{\arraystretch}{1.12}
\setlength{\litwidth}{\dimexpr\linewidth-18\tabcolsep\relax}
\begin{tabular}{@{}>{\raggedright\arraybackslash}p{.180\litwidth}>{\raggedright\arraybackslash}p{.110\litwidth}>{\raggedright\arraybackslash}p{.050\litwidth}>{\raggedright\arraybackslash}p{.085\litwidth}>{\raggedright\arraybackslash}p{.130\litwidth}>{\centering\arraybackslash}p{.080\litwidth}>{\centering\arraybackslash}p{.050\litwidth}>{\centering\arraybackslash}p{.045\litwidth}>{\centering\arraybackslash}p{.045\litwidth}>{\raggedright\arraybackslash}p{.225\litwidth}@{}}
\toprule
\multirow{2}{*}{\textbf{Study}} & \multirow{2}{*}{\textbf{Group}} & \multirow{2}{*}{\textbf{Age}} & \multirow{2}{*}{\textbf{Sample size}} & \multirow{2}{*}{\textbf{Region}} & \multirow{2}{*}{\textbf{Modeling}} & \multicolumn{3}{c}{\textbf{Model information}} & \multirow{2}{*}{\textbf{Usage}} \\
\cmidrule(lr){7-9}
& & & & & & \textbf{Shape} & \textbf{Size} & \textbf{Pose} & \\
\midrule
Wang et al. (2017)~\litcite{wang2017finite} & Chinese Male & 25--30 & 200 & Neck & $\times$ & $\checkmark$ & $\checkmark$ & $\checkmark$ & Garment design \\[2pt]

FLAME (2017)~\litcite{li2017flame} & US/European & 18--65 & 3,800 & Head-Neck & $\checkmark$ & $\checkmark$ & $\times$ & $\checkmark$ & Animation \\[2pt]

LYHM (2020)~\litcite{dai2020statistical} & Mostly White & 1--89 & 1,212 & Head & $\checkmark$ & $\checkmark$ & $\times$ & $\times$ & Head generation  \\[2pt]

Wang and Gu (2021)~\litcite{wang2021neckshoulder} & Chinese Female & 18--25 & 188 & Neck-Shoulder & $\times$ & $\checkmark$ & $\checkmark$ & $\checkmark$ &Garment design, anthropometry  \\[2pt]

Zhang et al. (2022a)~\litcite{zhang2022neckshoulder} & Chinese Male & 18--25 & 174 & Neck-Shoulder & $\times$ & $\checkmark$ & $\checkmark$ & $\checkmark$ &Garment design, anthropometry \\[2pt]

Zhang et al. (2022b)~\litcite{zhang2022age} & Chinese & 18--84 & 1,496 & Head & $\checkmark$ & $\checkmark$ & $\checkmark$ & $\times$ &Wearable design, anthropometry \\[2pt]

HIFI3D++ (2022)~\litcite{chai2022realy} & Caucasian/Asian & - & 1,957 & Head & $\checkmark$ & $\checkmark$ & $\times$ & $\times$ &Reconstruction \\[2pt]

HACK (2023)~\litcite{zhang2023hack} & Multiethnic & 15--65 & 624+600 & Head-Neck & $\checkmark$ & $\checkmark$ & $\times$ & $\checkmark$ & Animation \\[2pt]

GNM (2026)~\litcite{ploumpis2026gnmhead} & Multiethnic & 15--84 & 5000+ & Head-Neck & $\checkmark$ & $\checkmark$ & $\times$ & $\checkmark$ & Animation, simulation \\[2pt]

\textbf{Ours} & Chinese adults & 20--70 & 200 & Head-neck-shoulder & $\checkmark$ & $\checkmark$ & $\checkmark$ & $\times$ & Wearable design, anthropometry \\[2pt]
\bottomrule
\end{tabular}
\end{table*}

Statistical shape modeling of articulated body regions faces a fundamental challenge. Scans collected from different subjects differ in pose as well as in anatomy, and these two sources of variation become entangled in the resulting shape space~\cite{danckaers2019posture}. Two streams of work have addressed this entanglement. The first builds pose-aware models that factor shape and pose into separate parameter spaces~\cite{anguelov2005scape, loper2015smpl, li2017flame}. Representative examples include SCAPE~\cite{anguelov2005scape} and SMPL~\cite{loper2015smpl} for whole-body modeling, and FLAME~\cite{li2017flame} for the head and face. The second normalizes pose as a preprocessing step, then builds the shape model on the aligned data~\cite{danckaers2019posture}. Generalized Procrustes analysis (GPA)~\cite{Gower1975GPA} and related superimposition methods from geometric morphometrics~\cite{mitteroecker2009gm} are typical choices. Posture normalization for ergonomic design must preserve individual surface geometry and physical size. In addition, Table~\ref{tab:hns_literature_review} shows that existing shape models mainly cover the head or head-neck region, with limited shoulder coverage for integrated anthropometry. The only head-neck dataset, HACK~\cite{zhang2023hack}, is oriented toward digital human animation, with limited posterior accuracy and insufficient coverage of the neck-shoulder area for ergonomic design. These methodological requirements and data limitations constrain the scientific design of head-neck-shoulder supportive products.

\begin{figure*}[t!]
  \centering
  \setlength\abovecaptionskip{0pt}
  \includegraphics[width=1.\linewidth]{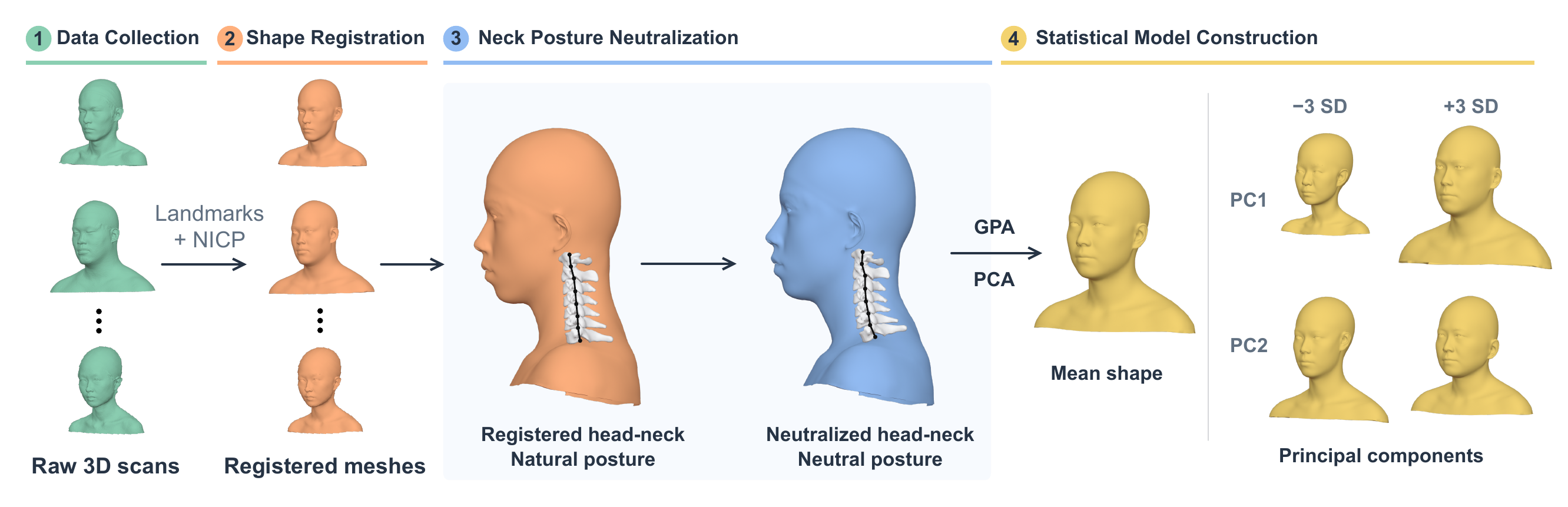}
  \caption{Overview of the head-neck-shoulder shape modelling pipeline. (1) Raw three-dimensional scans are collected in natural posture. (2) Anatomical landmarks and non-rigid iterative closest point (NICP) registration establish vertex correspondence. (3) Model-derived per-vertex transformations map each registered head-neck surface to a common neutral posture while retaining individual surface detail. (4) Generalized Procrustes analysis (GPA) aligns the meshes by rotation and translation. (5) Principal component analysis (PCA) represents population shape variation by the mean shape and principal components. The cervical bones~\cite{meng2020spine} and joint chains are shown for illustration only and do not represent experimentally measured skeletal data.}
  \label{fig:pipeline}
\end{figure*}

Motivated by this need, this paper develops a framework that couples 3D data acquisition, posture neutralization, and statistical shape modeling for the head-neck-shoulder region. We collect a scan from each of 200 Chinese adults (gender-balanced, aged 20-70 years) under a self-selected natural standing posture. To prevent inter-subject posture variation from contaminating the shape model, every scan is posture-neutralized before statistical modeling. We fit a parametric head-neck model to each scan to separate cervical pose from anatomy, and reset the cervical pose to a neutral configuration that reflects a relaxed, healthy posture. This age range covers the working-age and early-retirement population most affected by prolonged screen use and cervical degeneration~\cite{intaruk2025differences,zhuo2026incidence}. The Chinese cohort additionally supports population-specific sizing, given the reported differences in head morphology across populations~\cite{luximon2012chinese}. 
The main contributions of this work are summarized as follows:
\begin{itemize}
    \item We develop a statistical shape model (SSM) based on posture-neutralized scans, which captures genuine anatomical diversity rather than posture-induced deformation, providing a standardized shape space and population-level modes of variation to support parametric ergonomic design and population coverage analysis.

    \item We construct a novel 3D head-neck-shoulder dataset comprising 200 Chinese adults (balanced by gender, aged 20--70 years), addressing the critical lack of integrated head-neck-shoulder morphological data for ergonomic design.

     \item We propose a posture-neutralization pipeline for 3D head-neck-shoulder shape modeling. A parametric head-neck model is fitted to each scan to separate cervical pose from craniocervical anatomy, the cervical pose is reset to a common neutral configuration, and the neutralized head-neck is recombined with the subject's shoulder.
    
\begin{figure*}[t!]
  \centering
  \setlength\abovecaptionskip{0pt}
  \includegraphics[width=\linewidth]{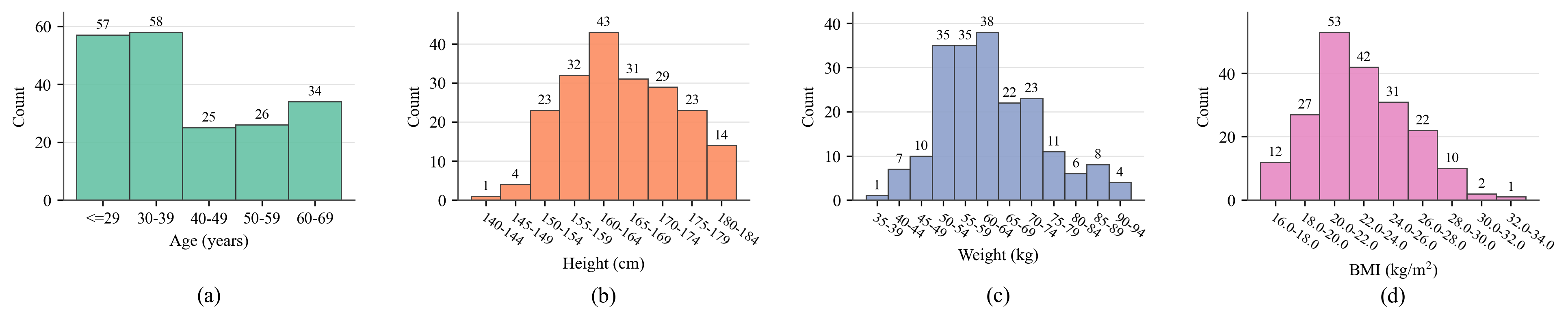}
  \caption{Distributions of participant characteristics in the study cohort ($n=200$). (a) Age. (b) Height. (c) Weight. (d) Body mass index (BMI).}
  \label{fig:participant_distributions}
\end{figure*}

\end{itemize}

\section{Method}

\subsection{Overview}

Figure~\ref{fig:pipeline} summarizes the data states used in the analysis. Registration first converts each raw scan to a mesh with shared topology and vertex correspondence while retaining the scan coordinate system. HACK fitting then estimates the subject-specific shape and acquisition posture. This change is applied directly to the registered head-neck vertices. The result is joined to the registered shoulder to form a complete head-neck-shoulder mesh. Generalized Procrustes analysis removes translation and rotation from these meshes, after which principal component analysis is applied to the resulting coordinate matrix.

\subsection{Data Collection}

Two hundred Chinese adults (100 male, 100 female) aged 20-70 years were recruited for the study. The distributions of age, height, weight, and body mass index (BMI) are shown in Fig.~\ref{fig:participant_distributions}. Subjects were eligible if they had no clinically diagnosed musculoskeletal disorders, no history of cervical injury or surgery, no congenital head or neck deformities, and no scoliosis or severe kyphosis. They were also required to stand independently in a stable upright posture for the duration of the scan. Subjects with metallic implants in the head or neck region were excluded, since these may interfere with optical scanning. This research complied with the American Psychological Association Code of Ethics. The study was approved by the Departmental Research Committee of The Hong Kong Polytechnic University (Reference Number: HSEARS20240912010). Written informed consent was obtained from each participant. Scanning took place in a private room with approximately $3~\text{m} \times 3~\text{m}$ of clear floor space, uniform overhead lighting, and a flat non-reflective floor. Geometry was acquired with a handheld Creality CR-Scan Otter together with the CreatyScan acquisition software, calibrated against a reference board at the start of each session. One trained operators handled calibration, posture guidance, and scanning throughout the study.

\textbf{Questionnaire.} Each subject completed a structured baseline questionnaire prior to scanning. The questionnaire covered ten categories: demographics, lifestyle and work habits, cervical history and health background, neck pain on a 0-10 numeric rating scale (VAS) \cite{koo2025visual}, the Neck Disability Index (NDI) \cite{vernon1991neck}, the Pittsburgh Sleep Quality Index (PSQI) \cite{buysse1989pittsburgh}, pillow use and preferences, forward-head-posture self-assessment, the PHQ-4 mental health screen, and overall self-rated health. These items are retained as subject-level metadata for later subgroup and correlation analyses.

\textbf{Scanning.} After completing the questionnaire, each subject changed into a standard scanning garment with an open neckline that exposed the clavicles and upper chest, and tucked all hair into an elastic cap so that the cranial vault remained visible. Reflective marker stickers were applied at anatomical landmarks on the neck and shoulders, corresponding to the body landmarks defined in Section~\ref{sec:landmarks}. The subject then adopted a self-selected natural standing posture, with feet shoulder-width apart, arms relaxed at the sides, gaze directed toward an eye-level target, and the head and neck held comfortably rather than forced to a fixed cervical angle. A reference image illustrating the standing protocol was posted at the scanning station as a visual cue. The scanner was configured for human-body geometry capture at the high-quality setting, and the operator acquired a single full scan by walking once around the subject. Each scan was inspected immediately for coverage, surface noise from clothing folds or stray hair, and motion artefacts. Scans that failed this check were redone on the spot. The procedure produced one textured triangular mesh per subject, for a total of 200 scans that serve as the input to the registration pipeline.

\begin{figure*}[p]
  \vspace{-55pt}
  \centering
  \captionsetup{skip=0pt}  
  \includegraphics[width=.95\linewidth]{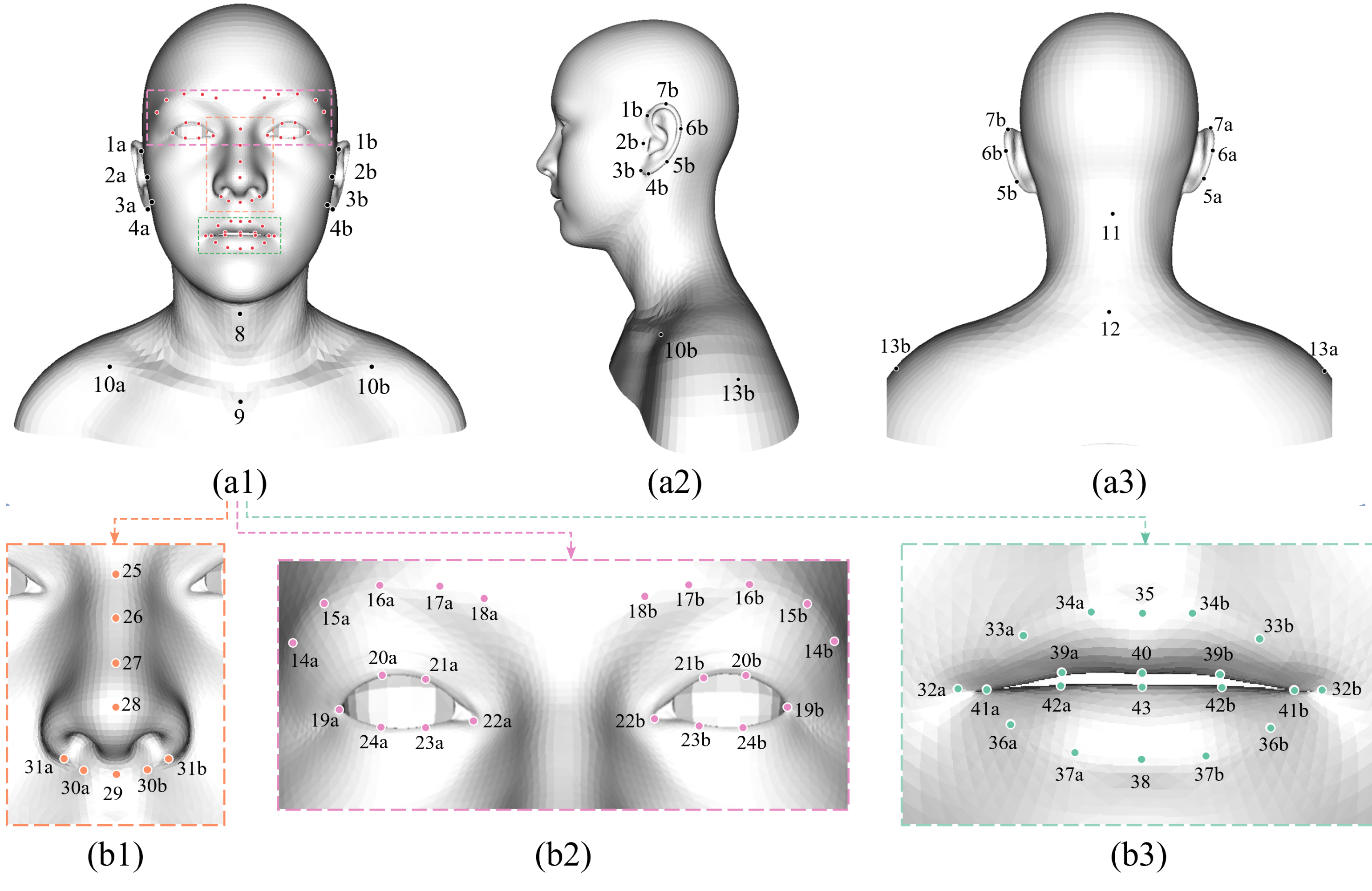}
  \caption{Our predefined $73$ landmarks on the template mesh, including 22 ear-neck-shoulder landmarks and 51 facial landmarks. (a) Ear-Neck-Shoulder landmarks (1-13): shown in front (a1), side (a2), and back (a3) views. Indices 1-4 are bilateral pairs along the anterior helix (1, helix apex, 2, upper helix, 3, lower anterior helix, 4, pre-auricular notch), and 5-7 are bilateral pairs along the posterior helix (5, posterior helix near antitragus, 6, mid posterior helix, 7, posterior helix apex). Midline points 8 and 9 are the thyroid notch and the suprasternal (jugular) notch. Indices 10a/10b are the supraclavicular fossae. Posterior midline points 11 and 12 are the high cervical point near the C1 level and the C7 vertebra prominens. Indices 13a/13b are the acromion tips. (b) Facial landmarks (14-43) are shown in three zoomed panels. The nose (b1) includes points 25-29 along the midline from nasion to subnasale, and pairs 30a/30b and 31a/31b at the inner and outer nostril edges. The eyebrows and eyes (b2) include brow points 14-18 in five bilateral pairs from lateral to medial, and eye points 19-24 in six pairs running from outer canthus around the eye to lower lid lateral. The mouth (b3) includes outer lip points 32-38 from mouth corner to lower-lip midline, and inner lip points 39-43 from paramedian to midline. Facial indices follow the iBUG 300-W convention~\cite{Sagonas2016iBUG}, reusing points 17-67 of the 68-point scheme. Throughout, suffixes \texttt{a}/\texttt{b} denote the left/right side of the image.}
  \label{fig:landmarks}

  \captionsetup{skip=0pt}  
  \includegraphics[width=.95\linewidth]{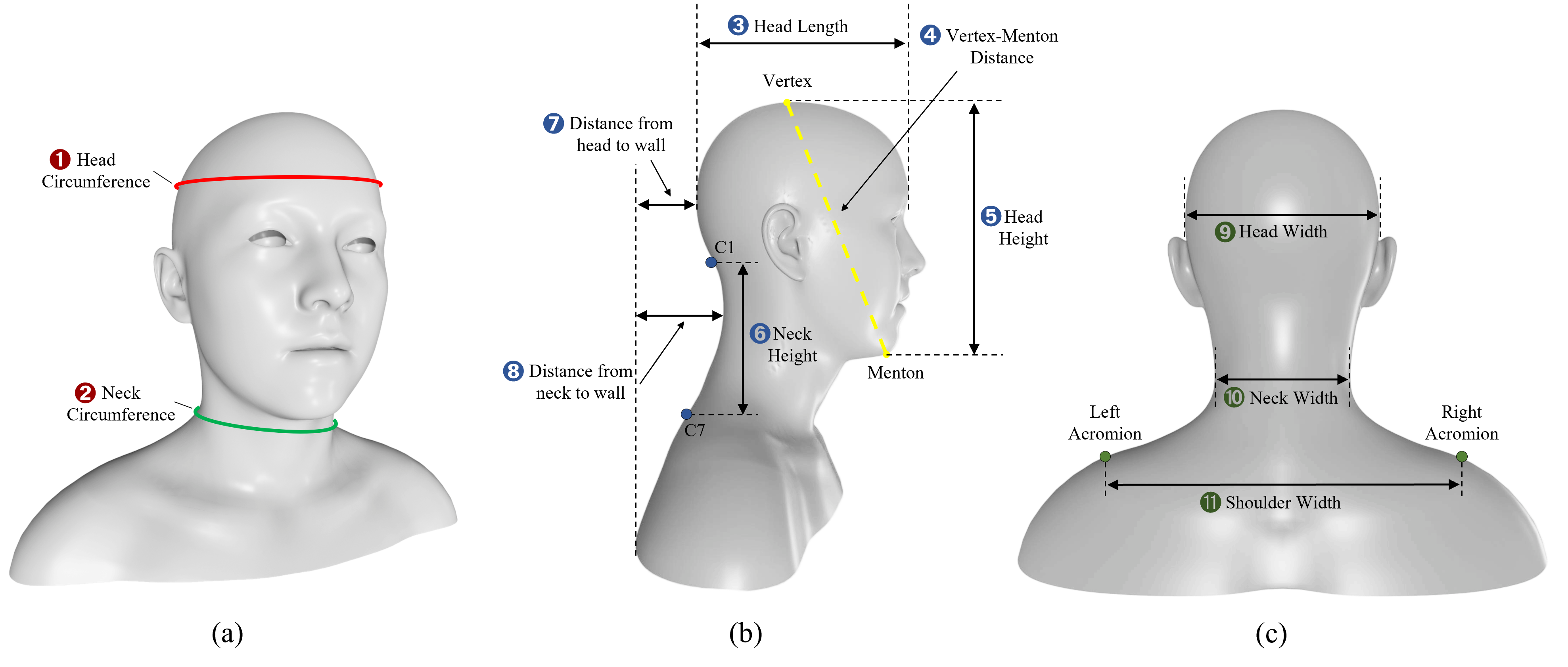}
  \caption{Anthropometric measurements illustrated on a representative subject from three viewpoints; the circled numbers index the eleven measurements. (a) Frontal view: \dnum{1} the head circumference (red contour) at the maximum cranial girth, and \dnum{2} the neck circumference (green contour) at the level of the thyroid cartilage, each obtained as the perimeter of the polygon formed by intersecting the mesh with the corresponding horizontal plane. (b) Lateral view: \dnum{3} the head length, the horizontal distance from the forehead to the back of the head; \dnum{4} the vertex-menton distance (yellow dashed line), the straight-line distance from the vertex at the top of the head to the menton at the tip of the chin; \dnum{5} the head height, the vertical distance between the vertex and the menton; \dnum{6} the neck height, the vertical distance between the C1 and C7 cervical levels; \dnum{7} the head-to-wall distance, the shortest distance from the posterior head to the vertical reference (wall) plane; and \dnum{8} the neck-to-wall distance, the greatest distance from the posterior neck to the same plane. (c) Posterior view: \dnum{9} the head width, \dnum{10} the neck width, and \dnum{11} the shoulder width (the left-to-right distance between the acromia), each a horizontal distance between bilateral landmarks.}
  \label{fig:anthropometric}
\end{figure*}

\subsection{Landmark Annotation}
\label{sec:landmarks}

Landmarks act as the common reference between scans and the template throughout the rest of the pipeline. They provide the correspondences that drive the rigid alignment, anchor the non-rigid registration that follows, define the C7 and tragus points used for cervical posture neutralization, and supply the geometric reference for the anthropometric measurements. To support all of these uses with a single annotation pass, we define a fixed set of $N_L = 73$ landmarks on the template mesh (see Fig.~\ref{fig:landmarks}), comprising $22$ body landmarks (1-13) and $51$ facial landmarks (14-43). Each landmark on the template is paired with a corresponding annotation on every scan. Midline points are labelled with a single integer (8, 9, 11, 12, 25-29, 35, 38, 40, 43), and bilateral pairs share an integer with the suffix \texttt{a}/\texttt{b} denoting the left/right side of the image rather than the anatomical side, so that the same numbering can be read consistently across front, side, and back views. Fig.~\ref{fig:landmarks}(a) shows the body landmarks in three views, and Fig.~\ref{fig:landmarks}(b) shows the facial landmarks in three zoomed panels covering the nose (b1), eyebrows and eyes (b2), and mouth (b3).

\textbf{Ear-Neck-Shoulder Landmarks.} The $22$ body landmarks are organized by anatomical region. Indices 1-4 cover the anterior helix from the helix apex down to the pre-auricular notch on each side, and indices 5-7 cover the posterior helix from the antitragus up to the posterior helix apex. The anterior cervical midline is anchored by the thyroid notch (8) and the suprasternal notch (9), and the posterior cervical midline by a high cervical point near the C1 level (11) and the C7 vertebra prominens (12). Indices 10a/10b mark the supraclavicular fossae, and 13a/13b mark the acromion tips. The cervical midline points (8, 9, 11, 12) and the shoulder line (10, 13) are deliberately densified relative to head-only templates, because the cervical and shoulder geometry is the focus of the present model.

\textbf{Facial Landmarks.} The $51$ facial landmarks follow the iBUG 300-W convention~\cite{Sagonas2016iBUG}, reusing indices 17-67 of the 68-point scheme and excluding the 17-point jaw contour. The five facial groups are eyebrow (14-18, five pairs), eye (19-24, six pairs), nose (25-31, five midline points and two pairs along the nasal alae), outer lip (32-38, five pairs and two midline points), and inner lip (39-43, three pairs and two midline points). For each scan, these $51$ points are detected automatically rather than placed by hand. The textured mesh is rendered from three viewpoints, and a face alignment network~\cite{Bulat2017FaceAlignment,Kazemi2014FaceAlignment} predicts 2D landmarks in each rendered image. Each 2D detection is then back-projected onto the mesh surface, and detections of the same landmark from different views are reconciled by selecting the candidate with the smallest reprojection residual. Automatic detection is essential at the scale of this study, where manual placement of dense facial points across $200$ subjects would be neither consistent nor practical.

\subsection{Anthropometric Measurements}
\label{sec:anthropometric_measurements}

We compute eleven anthropometric measurements on the head-neck-shoulder mesh of each subject, illustrated from three viewpoints in Fig~\ref{fig:anthropometric}. They fall into four geometric types: two circumferences, six axis-aligned linear dimensions, one Euclidean chord, and two posture-related wall distances. Each mesh is expressed in a common anatomical frame with unit axes $\hat{\mathbf{x}}$ (medio-lateral), $\hat{\mathbf{y}}$ (vertical, superior-inferior), and $\hat{\mathbf{z}}$ (antero-posterior). All point coordinates and unit axes in this subsection are column vectors in $\mathbb{R}^{3 \times 1}$. These quantities provide a per-subject anthropometric summary for population statistics and serve to validate the posture normalization.

\textbf{Circumferences.} The two circumferences (Fig.~\ref{fig:anthropometric}a) are obtained by intersecting the mesh with a plane defined by three anatomical points on the skin. For the head circumference (\dnum{1}), the three points lie on the maximum horizontal cranial girth: one at the forehead just above the supraorbital ridge, and two on the posterior cranium superior to the ears, following the convention of ISO~7250-1~\cite{ISO7250}. For the neck circumference (\dnum{2}), the three points lie at the level of the thyroid cartilage, two on the lateral sides of the neck and one on the anterior midline. Each circumference is then the perimeter of the closed polygon formed by intersecting the mesh $\mathcal{S}$ with the corresponding plane $\Pi$:
\begin{equation}
\label{eq:circumference}
C = \sum_{i=1}^{K} \|\mathbf{c}_{i+1} - \mathbf{c}_i\|, \quad \mathbf{c}_{K+1} \equiv \mathbf{c}_1,
\end{equation}
where $\{\mathbf{c}_i\}_{i=1}^{K}$ are the ordered vertices of $\Pi \cap \mathcal{S}$, obtained by linear interpolation at each triangle edge that $\Pi$ crosses, and $\|\cdot\|$ is the Euclidean norm.

\textbf{Sagittal lengths and heights.} The lateral view (Fig.~\ref{fig:anthropometric}b) provides four linear dimensions. We write each axis-aligned dimension as the absolute projection of the vector between two points onto a unit anatomical axis,
\begin{equation}
\label{eq:projection}
D = \left| (\mathbf{v}_a - \mathbf{v}_b) \cdot \hat{\mathbf{e}} \right|, \qquad \hat{\mathbf{e}} \in \{\hat{\mathbf{x}}, \hat{\mathbf{y}}, \hat{\mathbf{z}}\}.
\end{equation}
The \emph{head length} (\dnum{3}) is the maximum cranial length, the antero-posterior extent of the head from the glabella to the opisthocranion, the most posterior point of the cranium~\cite{ISO7250}; it is Eq.~\eqref{eq:projection} with $\hat{\mathbf{e}} = \hat{\mathbf{z}}$. The \emph{vertex--menton distance} (\dnum{4}) is instead the full Euclidean chord between the vertex, the highest point of the head, and the menton, the lowest point of the chin,
\begin{equation}
\label{eq:vertexmenton}
L = \| \mathbf{v}_{\text{vertex}} - \mathbf{v}_{\text{menton}} \|,
\end{equation}
and the \emph{head height} (\dnum{5}) is its vertical component, given by Eq.~\eqref{eq:projection} for the same two points with $\hat{\mathbf{e}} = \hat{\mathbf{y}}$. The \emph{neck height} (\dnum{6}) applies Eq.~\eqref{eq:projection} with $\hat{\mathbf{e}} = \hat{\mathbf{y}}$ to the high cervical point near the C1 level and the C7 vertebra prominens.

\textbf{Wall distances.} The remaining two lateral measures quantify sagittal posture relative to the upper back. We define the wall plane $\Pi_w$ as the vertical plane passing through the scapulae, with unit normal $\hat{\mathbf{z}}$; the distance from a surface point $\mathbf{v}$ to this plane is,
\begin{equation}
\label{eq:wall}
d(\mathbf{v}, \Pi_w) = \left| (\mathbf{v} - \mathbf{q}) \cdot \hat{\mathbf{z}} \right|, \qquad \mathbf{q} \in \Pi_w.
\end{equation}
The \emph{head-to-wall distance} (\dnum{7}) is the minimum of $d$ over the posterior head, and the \emph{neck-to-wall distance} (\dnum{8}) is the maximum of $d$ over the posterior neck. They capture the depth of the hindbrain and the cervical curvature.

\textbf{Widths.} The posterior view (Fig.~\ref{fig:anthropometric}c) provides three medio-lateral widths, each obtained from Eq.~\eqref{eq:projection} with $\hat{\mathbf{e}} = \hat{\mathbf{x}}$. The \emph{head width} (\dnum{9}) is the maximum transverse diameter of the head, taken between the two euryons, the most lateral points of the cranium~\cite{ISO7250}; the \emph{neck width} (\dnum{10}) is taken at the narrowest lateral extent of the neck; and the \emph{shoulder width} (\dnum{11}) is the biacromial distance between the left and right acromia.

\begin{figure*}[t!]
  \centering
  	\captionsetup{skip=0pt}  
  \includegraphics[width=.95\linewidth]{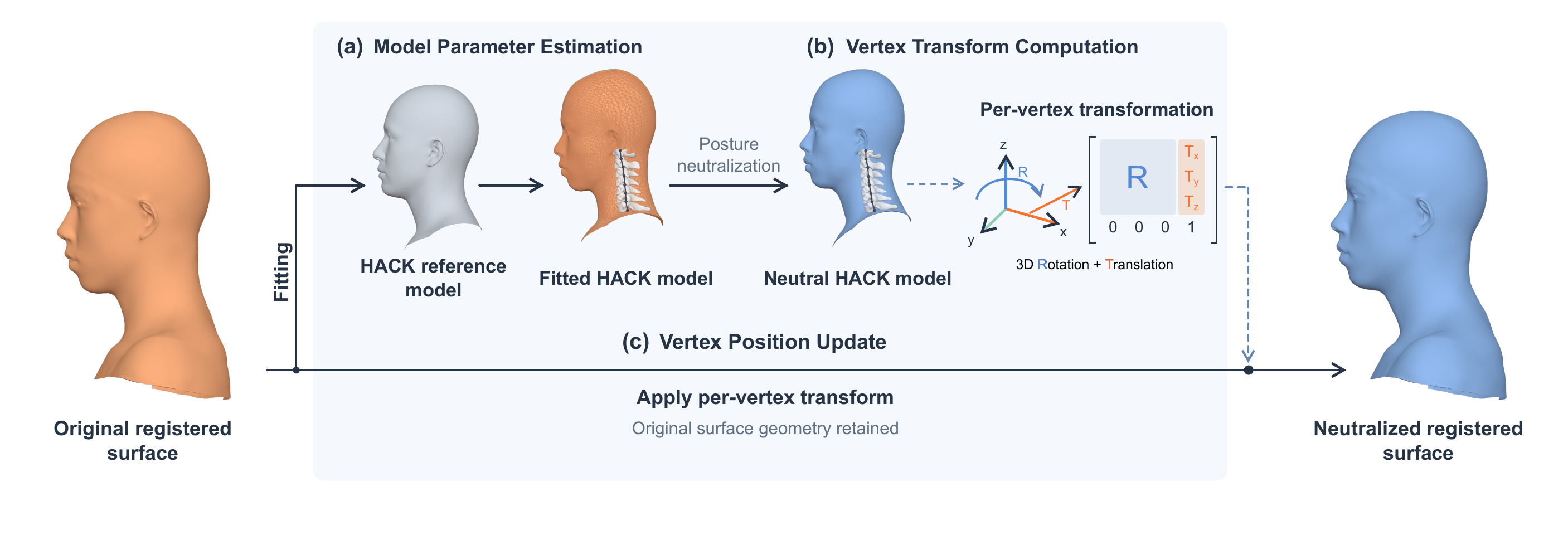}
  \caption{Neck posture neutralization. (a) The HACK reference model is fitted to the registered head-neck surface to estimate individual shape, acquired posture and the global similarity transform between model and registered coordinates. (b) The fitted shape is held fixed while cervical posture is reset to the common neutral configuration. Joint transformations and fixed skinning weights define the per-vertex neutralization. $R$: Rotation Matrix $\in R^{3\times3}$. (c) The model-derived neutralization is transferred through vertex correspondence to update the original registered surface, with coordinate changes determined by the fitted similarity transform. The cervical bones and joint chains are shown for illustration only and do not represent experimentally measured skeletal data.}
  \label{fig:pose_neutralization}
\end{figure*}

\begin{figure*}[t!]
  \centering
  	\captionsetup{skip=0pt}  
  \includegraphics[width=1.\linewidth]{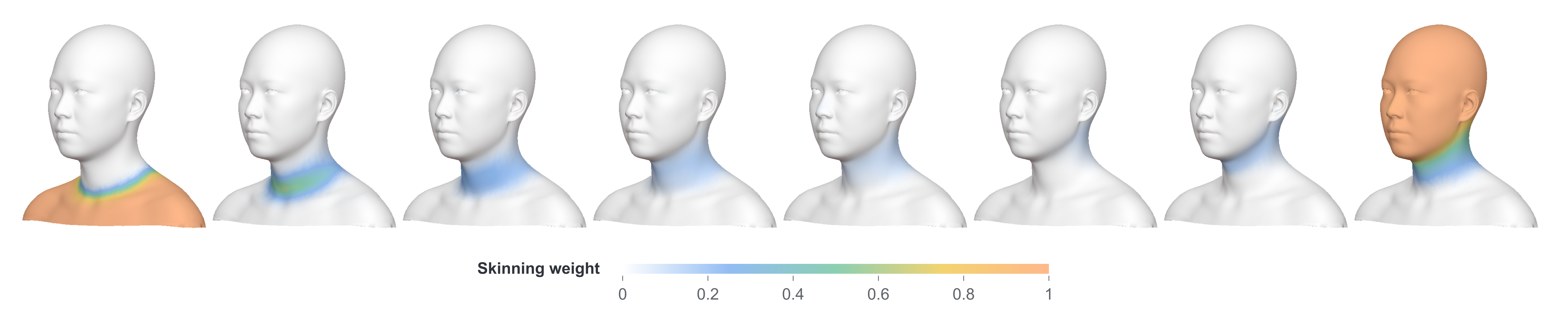}
  \caption{Cervical skinning weights on a representative neutralized registered head-neck surface. Weight fields for the C7–T1, C6–C7, C5–C6, C4–C5, C3–C4, C2–C3, C1–C2 and occiput–C1 (Oc–C1) joints, respectively. }
  \label{fig:weights}
\end{figure*}

\subsection{Shape Registration}

Shape registration establishes a common mesh representation across subjects. We deform the same head-neck-shoulder template onto every scan. The template and every registered output contain $n=21{,}212$ vertices with identical connectivity and vertex order. A vertex index therefore refers to the same surface location in all subjects. Point distribution models require such point-wise correspondence before population variation can be analysed~\cite{cootes1995active}. Template-based registration with fixed topology has also been used in statistical models of human head and craniofacial shape~\cite{booth2018large,dai2020statistical}. In our pipeline, the common correspondence supports the later posture neutralization, model alignment, and principal component analysis.

\textbf{Initial Alignment.} The raw scan and template can differ in position, orientation, and size. We use the 73 anatomical landmarks defined in Section~\ref{sec:landmarks} to obtain a landmark-guided global alignment before local deformation. The alignment begins with Procrustes-based similarity estimates from corresponding landmarks~\cite{Gower1975GPA}. A rigid ICP refinement then uses surface geometry to reduce the remaining mismatch in position and orientation~\cite{Besl1992ICP}. Together, these operations place the scan in the template working coordinate system. The complete scan-to-template similarity transform is recorded. Uniform scale is estimated during this initialization and is not updated during the subsequent NICP step.

\textbf{Non-rigid Deformation.} A single global transform cannot establish dense correspondence when local head-neck-shoulder geometry differs between the template and scan. We therefore apply NICP, which assigns a local affine transform to each template vertex and deforms the template toward the aligned scan~\cite{amberg2007optimal,allen2003space}. The optimization is guided by surface proximity, landmark correspondence, and deformation regularization. The regularization combines local stiffness with smoothing between neighboring vertices. Region-dependent surface weights and distance thresholds reduce the influence of unreliable correspondences in different anatomical regions. The global scale remains fixed during this step. After optimization, we apply the inverse of the recorded scan-to-template transform. The registered mesh for subject $i$ is denoted by $\mathcal{F}_i=\{\mathbf{f}_{i,v}\}_{v=1}^{n}$. Here, $\mathbf{f}_{i,v}\in\mathbb{R}^{3}$ represents vertex $v$ of this registered mesh and contains its three coordinates. The mesh has the template topology and vertex order, but its coordinates and physical scale are those of the subject's scan.

\subsection{Neck Posture Neutralization}
The registered surface contains anatomical variation, acquisition posture, and differences between the scan and the parametric model. We use HACK~\cite{zhang2023hack} to estimate the subject's shape and acquisition posture. The procedure has three steps. We first fit HACK to the registered head-neck surface, then transport the scan-to-model residual to a shared reference posture, and finally join the neutralized head-neck surface to the registered shoulder.

\subsubsection{Reference Head-Neck Model}
HACK represents the first $N_H=20{,}481$ vertices of each registered mesh~\cite{zhang2023hack}. These vertices cover the head and neck and have the same order as the HACK surface. The following equations describe the model for a single subject. The index $v\in\{1,\ldots,N_H\}$ denotes a vertex, and $b\in\{1,\ldots,8\}$ denotes a joint in the chain ($c_{7}t_{1}\!\rightarrow\!\cdots\!\rightarrow\!c_{1}c_{2}\!\rightarrow\!oc_{1}$). The fitted parameters are the shape vector $\beta\in\mathbb R^{200\times1}$ and the cervical-pose matrix $\theta\in\mathbb R^{8\times3}$. Each row of $\theta$ is a rotation vector in the local joint coordinate system. Its direction gives the rotation axis and its magnitude gives the angle in radians.

The shape-dependent point at vertex $v$ is denoted by $\boldsymbol{\gamma}_v(\beta)$. The pose-corrective offset at the same vertex is denoted by $\boldsymbol{\delta}_v(\theta)$. These are two independent functions of the shape and pose parameters, respectively.

The fixed cervical skeleton contains eight joints. For joint $b$, $\mathbf G_b(\theta) \in \mathbb R^{4\times4}$ is the transformation from the binding configuration to posture $\theta$. Each vertex has a scalar skinning weight $w_{v,b}\in[0,1]$, with $\sum_{b=1}^{8}w_{v,b}=1$. The weighted joint transforms $\mathbf M_v(\theta) \in \mathbb R^{4\times4}$ give
\begin{equation}
\label{eq:blend}
\mathbf M_v(\theta)
=\sum_{b=1}^{8}w_{v,b}\mathbf G_b(\theta)
\end{equation}
The complete HACK surface for one subject is then represented by
\begin{equation}
\label{eq:lbs}
\mathbf H(\beta,\theta)
=
\left[
\mathbf M_v(\theta)
\left(
\boldsymbol{\gamma}_v(\beta)+\boldsymbol{\delta}_v(\theta)
\right)
\right]_{v=1}^{N_H}
\end{equation}
The bracket collects the $N_H$ vertex columns. Thus, $\mathbf H(\beta,\theta)\in\mathbb R^{4\times N_H}$ directly denotes the complete posed HACK surface.

\subsubsection{Model Parameter Estimation} 
Before the neck posture can be neutralized, we need to determine the subject's shape and the posture adopted during scanning. We therefore fit HACK to the registered head-neck surface. The fitted parameters are $\beta$, $\theta$, and a similarity transform $\mathbf T\in\mathbb R^{4\times4}$. The transform $\mathbf T$ maps HACK model coordinates to registered-scan coordinates.

The registered head-neck surface is written as $\mathbf F^{hn}\in\mathbb R^{4\times N_H}$. The parameters are estimated by minimizing the scalar objective
\begin{equation}
\label{eq:hack_fit}
E=d\!\left(\mathbf T \cdot \mathbf H(\beta,\theta),\mathbf F^{hn}\right)
+\lambda_\beta\lVert\beta\rVert_2^2
+\lambda_\theta\lVert\theta\rVert_F^2.
\end{equation}
Here, $d(\cdot,\cdot)$ returns a scalar distance between two surfaces. The non-negative scalars $\lambda_\beta$ and $\lambda_\theta$ control the shape and pose constraints. The optimization returns $\beta^*$, $\theta^*$, and $\mathbf T$. The fitted shape $\beta^*$ is kept fixed during posture neutralization.

\subsubsection{Vertex Transformation Computation} 
After model fitting, we calculate the change that moves each head-neck vertex from the acquired posture to the common reference posture. We define the reference posture as $\theta_n=\mathbf 0_{8\times3}$, which sets all eight cervical rotation vectors to zero. For each head-neck vertex, the change from the fitted posture $\theta^*$ to the reference posture $\theta_n$ is
\begin{equation}
\label{eq:displace}
\mathbf D_v
=\mathbf M_v(\theta_n)\mathbf M_v(\theta^*)^{-1}
\end{equation}
The matrix $\mathbf D_v\in\mathbb R^{4\times4}$ is the fitted-to-reference posture transform for vertex $v$.

The per-vertex matrices $\mathbf D_v$ define a displacement field over the head-neck surface. We apply this field to the registered surface to obtain the neutralized vertex positions. Let $\mathbf f_v\in\mathbb R^{4\times1}$ be the registered head-neck vertex and $\mathbf y_v\in\mathbb R^{4\times1}$ its neutralized position. The updated vertex is computed as
\begin{equation}
\label{eq:update}
\mathbf y_v
=\mathbf T\mathbf D_v\mathbf T^{-1}\mathbf f_v.
\end{equation}
The inverse transform $\mathbf T^{-1}$ maps the registered vertex to HACK coordinates. The matrix $\mathbf D_v$ applies the fitted-to-reference posture change. The transform $\mathbf T$ then returns the vertex to registered-scan coordinates.

\subsection{Mesh Alignment}

Vertex correspondence alone does not remove differences in mesh position and orientation. We therefore align the complete head-neck-shoulder meshes before PCA using generalized Procrustes analysis (GPA)~\cite{Gower1975GPA}. GPA uses the first three Cartesian coordinates of each homogeneous vertex. The alignment is estimated from 1,993 neck vertices and then applied to the complete mesh. It removes rotation and translation but does not change scale or reflect the surface. 

After alignment, the coordinates of subject $i$ are flattened into the column vector $x_i\in\mathbb R^{3n\times1}$. For $m$ subjects, the PCA input matrix is
\begin{equation}
X=[x_1,x_2,\ldots,x_m]^{\top}\in\mathbb R^{m\times3n}.
\end{equation}

\subsection{Statistical Model Construction}

The final modelling step represents population-level shape variation in a lower-dimensional form. We apply principal component analysis to $X$ to construct the statistical shape model~\cite{jolliffe2016pca}. The model represents the population using a mean shape and principal components, with each eigenvalue giving the variance explained by its component.

Let $\bar{x}\in\mathbb{R}^{3n\times1}$ denote the mean shape vector. Let $P=[\mathbf{p}_1,\ldots,\mathbf{p}_k]\in\mathbb{R}^{3n\times k}$ contain the first $k$ principal directions, ordered by decreasing eigenvalue. Each column $\mathbf{p}_j\in\mathbb{R}^{3n\times1}$ is one principal direction. The corresponding eigenvalues form the diagonal matrix $\Lambda=\operatorname{diag}(\lambda_1,\ldots,\lambda_k)\in\mathbb{R}^{k\times k}$, where $\lambda_j$ is the non-negative eigenvalue of component $j$.

A shape vector $x\in\mathbb{R}^{3n\times1}$ is approximated by
\begin{equation}
\label{eq:ssm}
x=\bar{x}+P\alpha,
\end{equation}
where $\alpha=[\alpha_1,\ldots,\alpha_k]^{\top}\in\mathbb{R}^{k\times1}$ contains the PC scores. Each score gives the deviation from the mean along its corresponding principal direction.

\section{Results}

\subsection{Posture-Neutralized Statistical Shape Model}
\label{sec:neutralized_ssm_results}

\subsubsection{Average Head-Neck-Shoulder Shape}

\begin{figure}[t!]
  \centering
  \captionsetup{skip=0pt}  
  \includegraphics[width=\linewidth]{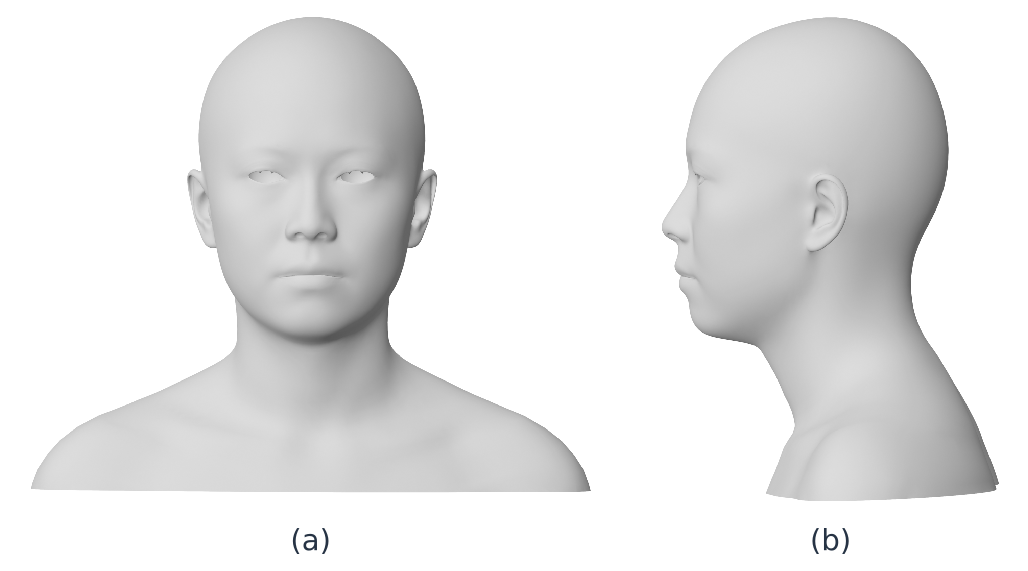}
  \caption{Average posture-neutralized head-neck-shoulder shape. The same mean mesh is shown in (a) frontal and (b) lateral views.}
  \label{fig:average_head_neck_shape}
\end{figure}

Figure~\ref{fig:average_head_neck_shape} shows the average posture-neutralized head-neck-shoulder shape in frontal and lateral views. This mean shape provides the reference for the principal component variations reported below.

\begin{figure*}[t!]
  \centering
  	\captionsetup{skip=0pt}  
  \includegraphics[width=\textwidth]{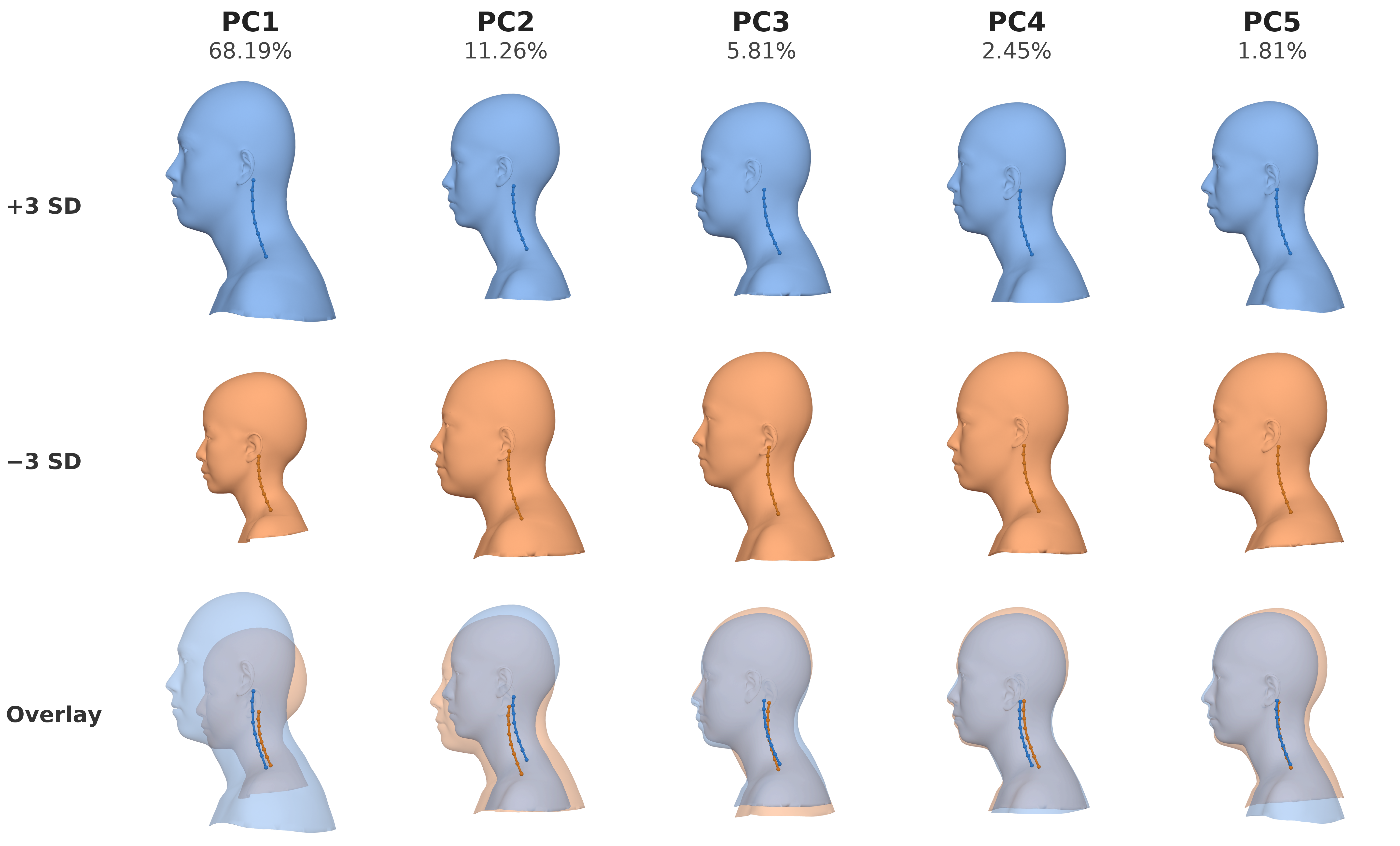}
  \caption{First five principal components of the posture-neutralized statistical shape model. The first and second rows show the $+3$ SD and $-3$ SD endpoints in the left lateral view. The third row overlays both endpoints. Slate blue denotes $+3$ SD, and orange denotes $-3$ SD.}
  \label{fig:neutralized_pc_modes}
 	\captionsetup{skip=0pt}  
  \includegraphics[width=1.\textwidth]{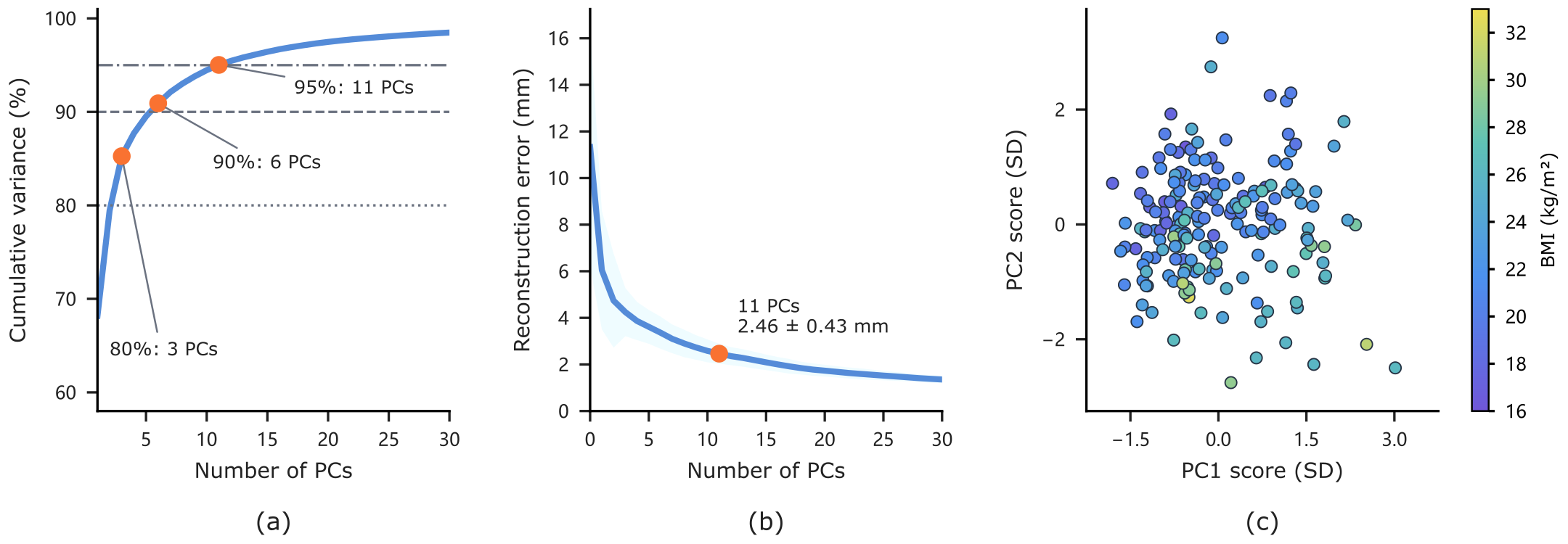}
  \caption{Evaluation and analysis of posture-neutralized head-neck shoulder model SSM. (a) Compactness: cumulative explained variance of the first 30 principal components. (b) Compactness: mean reconstruction error with standard deviation for different numbers of principal components. (c) BMI distribution of the standardized first and second PC scores in the dataset.}
  \label{fig:neutralized_ssm_evaluation}
\end{figure*}

\subsubsection{Principal Components and Model Evaluation}
\paragraph{M\textbf{odel Visualization}}
Figure~\ref{fig:neutralized_pc_modes} shows the first five principal components of the posture-neutralized statistical shape model. Together, these components explained 89.52\% of the total shape variance. PC1 explained 68.19\%, followed by PC2 at 11.26\%, PC3 at 5.81\%, PC4 at 2.45\%, and PC5 at 1.81\%. Each component is shown at $-3$ SD and $+3$ SD from the mean shape.

\paragraph{\textbf{Model Evaluation}} Figure~\ref{fig:neutralized_ssm_evaluation} summarizes the variance structure and training-set reconstruction performance of the model fitted to 200 aligned meshes. Three principal components were required to explain 80\% of the variance, 6 were required to explain 90\%, and 11 were required to explain 95\%. The mean per-vertex reconstruction error decreased from $11.331 \pm 5.156$ mm for the mean-shape baseline to $2.462 \pm 0.426$ mm when 11 principal components were retained.

\begin{figure*}[t]
  \centering
  \captionsetup{skip=0pt}  
  \includegraphics[width=1.\textwidth]{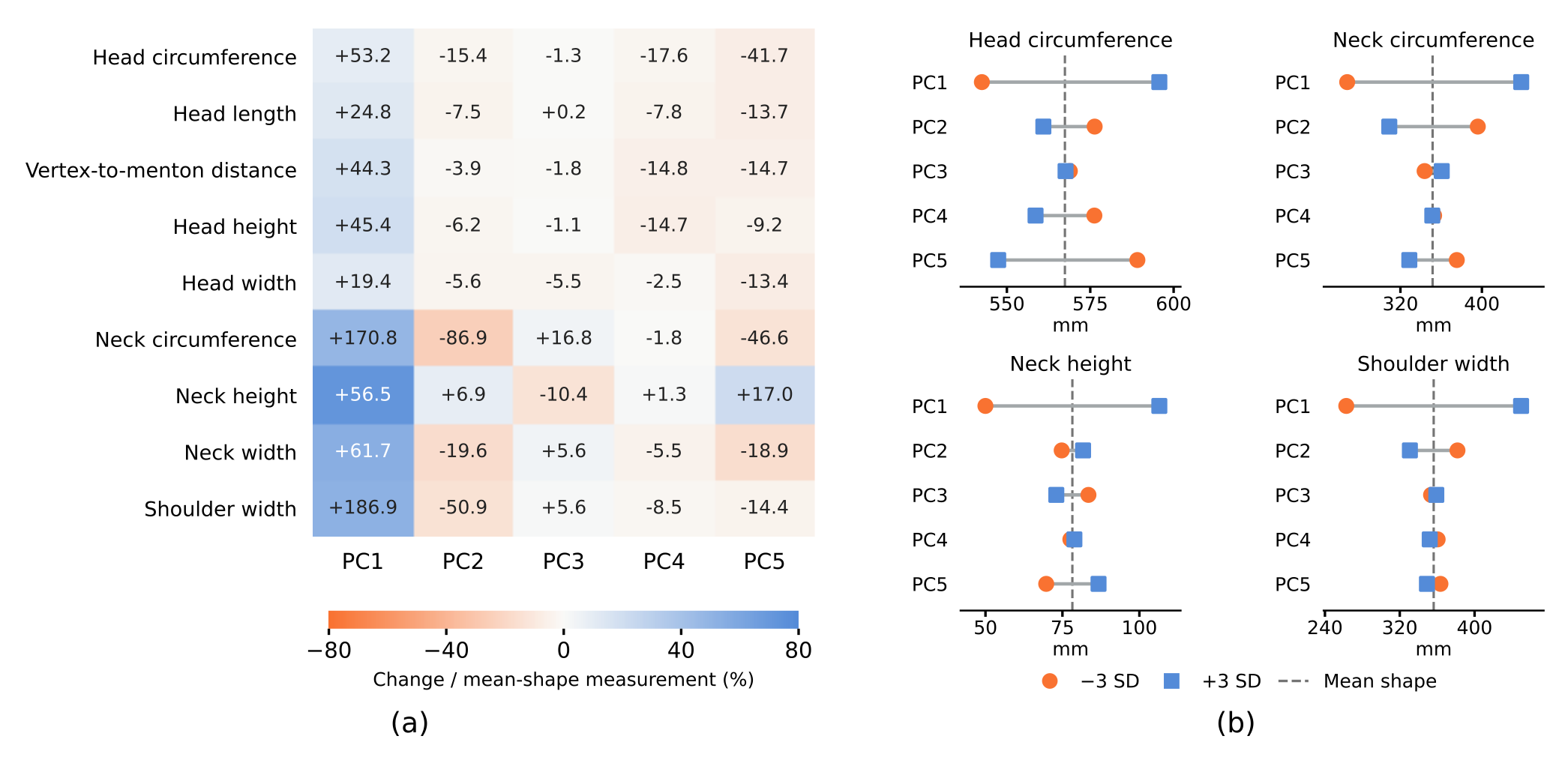}
  \caption{Anthropometric changes along the first five principal components of the posture-neutralized model. (a) Endpoint changes in nine dimensions. (b) Absolute endpoint measurements for head circumference, neck circumference, neck height, and shoulder width.}
  \label{fig:neutralized_pc_measurements}
\end{figure*}

\paragraph{\textbf{Model Analysis}} 
We interpreted the first five PCs using the size measurements defined in Section~\ref{sec:anthropometric_measurements} and shown in Fig.~\ref{fig:neutralized_pc_measurements}. All changes below refer to the model-generated endpoints from $-3$ SD to $+3$ SD. PC1 represented coordinated increases in head, neck, and shoulder dimensions. Head circumference, neck circumference, neck height, and shoulder width increased by 53.2, 170.8, 56.5, and 186.9~mm, respectively. PC2 mainly combined reductions in neck circumference and neck width with reduced shoulder width. These measurements decreased by 86.9, 19.6, and 50.9~mm, respectively, while neck height increased by 6.9~mm.

PC3 mainly changed neck height, which decreased by 10.4~mm. Neck circumference and neck width increased by 16.8 and 5.6~mm, respectively. PC4 mainly changed head height and vertex-to-menton distance, which decreased by 14.7 and 14.8~mm, respectively. PC5 combined increased neck height with reduced head and neck dimensions. Neck height increased by 17.0~mm, while neck circumference and neck width decreased by 46.6 and 18.9~mm, respectively. Head circumference, head length, and head width decreased by 41.7, 13.7, and 13.4~mm, respectively.

\begin{figure*}[!tp]
  \centering
  \captionsetup{skip=0pt}  
  \includegraphics[width=\textwidth]{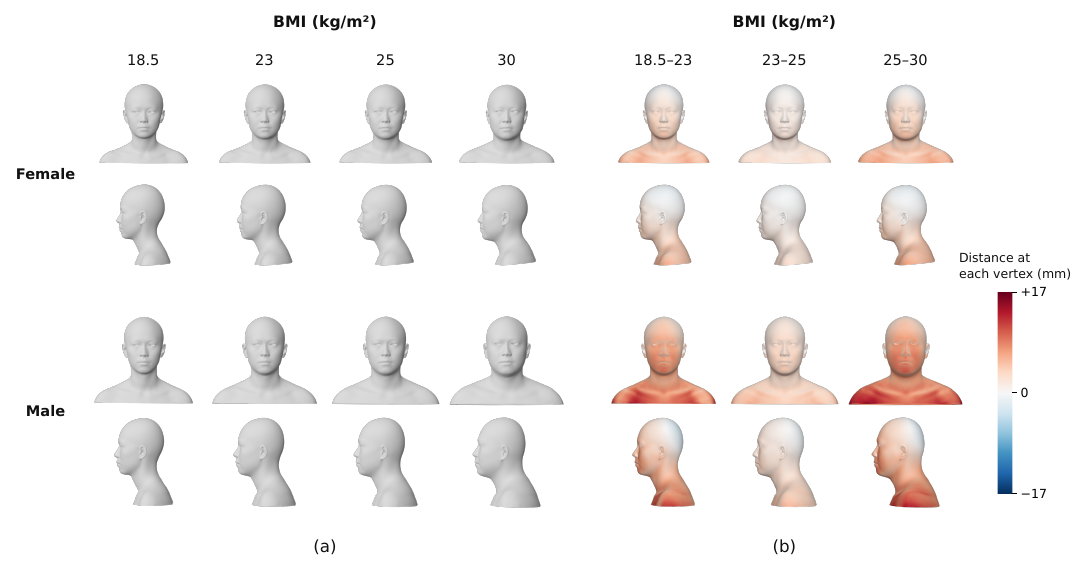}
  \caption{Predicted head-neck shape across gender and BMI. (a) Mean surfaces from gender-specific BMI regressions at BMI 18.5, 23, 25 and 30~kg/m$^2$. (b) Surface displacement between successive BMI levels. Red and blue indicate outward and inward normal displacement, respectively.}
  \label{fig:sex_bmi_models}
\end{figure*}

We used Pearson correlations to assess BMI associations with the first 11 PC scores separately in females and males. Gender-specific BMI regressions in the shared posture-neutralized PCA model predicted mean surfaces at BMI 18.5, 23, 25 and 30~kg/m$^2$. Figure~\ref{fig:sex_bmi_models} shows predicted mean shapes from the gender-specific BMI regressions. BMI was positively correlated with PC1 in females ($r=0.243$) and males ($r=0.516$). The corresponding correlations with PC2 were negative ($r=-0.415$ and $r=-0.582$, respectively). Higher BMI was associated with outward changes around the lower face, neck and shoulder region in these predicted surfaces.

\subsection{Model Comparisons}
\subsubsection{Neutralization vs.\ Non-neutralization}
Without posture neutralization, the first five PCs mixed size variation with changes in cervical posture. The model was fitted to registered surfaces from the same 200 participants, and these components explained 88.68\% of the total variance. PC1 mainly represented overall size. PC2, PC3 and PC5 combined changes in head and neck shape with cervical bending.

\begin{figure*}[t!]
  \centering
  \captionsetup{skip=0pt}  
  \includegraphics[width=\textwidth]{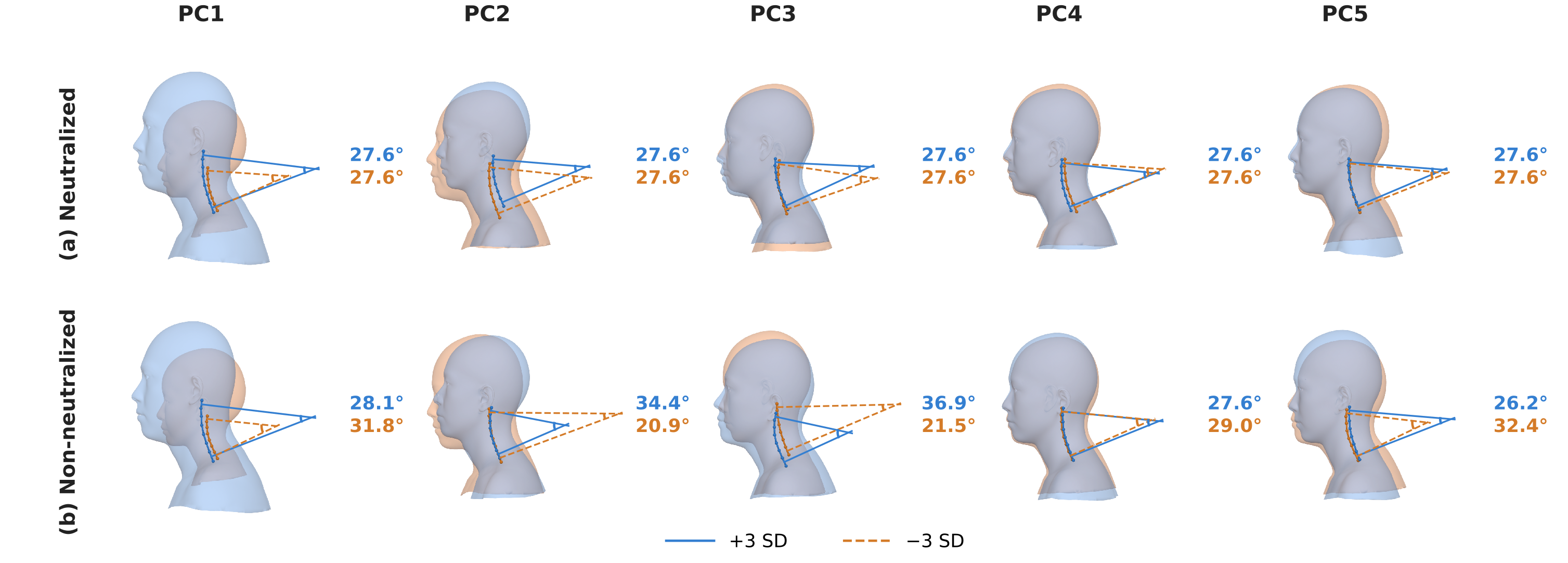}
  \caption{Enlarged cervical-chain Cobb angle comparison for the first five PC endpoints. Rows show (a) the neutralized model imposed and (b) the non-neutralized model with natural poses. Blue solid chains denote $+3$ SD, and orange dashed chains denote $-3$ SD.}
  \label{fig:cervical_chain_detail}
	\captionsetup{skip=0pt}  
  \includegraphics[width=\textwidth]{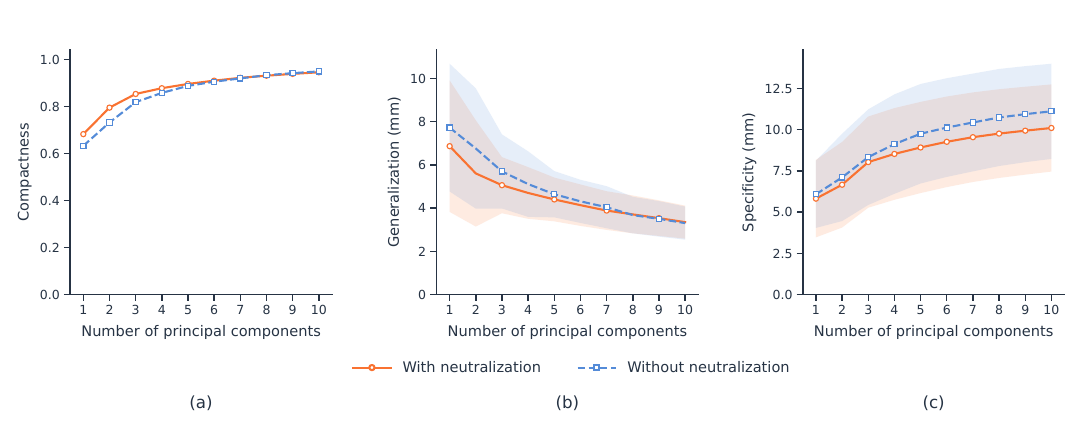}
  \caption{SSM evaluation with and without posture neutralization for the same participants. (a) Compactness: individual PC variance fractions and cumulative coordinate variance for the first ten components. (b) Generalization: mean corresponding-vertex RMSE on held-out surfaces using a shared five-fold partition. (c) Specificity: mean nearest-training-shape RMSE for generated shapes per dimension.}
  \label{fig:ssm_neutralization_metrics}
\end{figure*}

The enlarged cervical chains quantified this posture variation at the generated endpoints (Fig.~\ref{fig:cervical_chain_detail}). Without neutralization, the absolute HACK sagittal bone-axis angle differences were $13.50^{\circ}$, $15.38^{\circ}$ and $6.14^{\circ}$ for PC2, PC3 and PC5, respectively. The neutralized chains shared the imposed angle of $27.59^{\circ}$. The changes within the non-neutralized PCs show that shape variation cannot be interpreted independently of posture in this model.

\paragraph{\textbf{SSM Comparison}}
Posture neutralization concentrated more variance in the first components (Fig.~\ref{fig:ssm_neutralization_metrics}a). PC1 explained 68.19\% of coordinate variance after neutralization and 63.22\% without neutralization. The first five components explained 89.52\% and 88.68\%, respectively.

Generalization error decreased with the number of retained components (Fig.~\ref{fig:ssm_neutralization_metrics}b). At five components, RMSE was $4.404 \pm 1.013$~mm with neutralization and $4.645 \pm 1.071$~mm without neutralization. At ten components, the errors were $3.353 \pm 0.764$ and $3.308 \pm 0.767$~mm, respectively.

Specificity error was lower after neutralization across the first ten components (Fig.~\ref{fig:ssm_neutralization_metrics}c). At ten components, it was $10.107 \pm 2.643$~mm with neutralization and $11.118 \pm 2.888$~mm without neutralization.

\subsubsection{Model Construction Methods}
\label{sec:ssm_method_results}

We construct a PCA-based SSM that represents shape variation along the leading eigenvectors of the coordinate covariance matrix~\cite{cootes1995active,jolliffe2016pca}. We compare this representation with three alternatives. ICA seeks statistically independent components rather than directions of maximum variance~\cite{hyvarinen1999fastica}. Polynomial-kernel KPCA uses polynomial feature interactions to represent nonlinear variation, whereas RBF-kernel KPCA uses Gaussian similarities between shape vectors~\cite{scholkopf1998nonlinear}. These baselines compare linear variance-based, linear independence-based, and nonlinear kernel-based representations of shape variation.

\begin{figure*}[htp]
  \centering
  \captionsetup{skip=0pt}  
  \includegraphics[width=\textwidth]{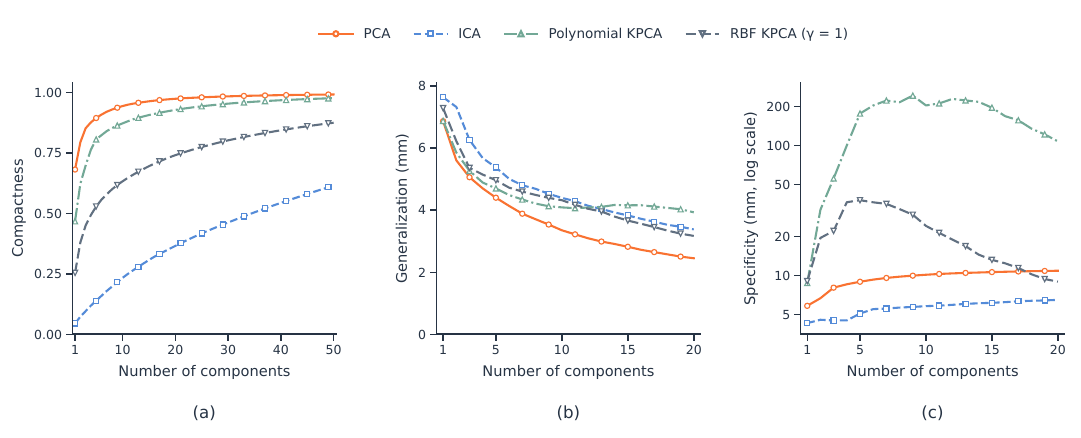}
  \caption{Evaluation of four SSM representations. (a) Compactness for the first 50 components of the full dataset. (b) Generalization for the first 20 components. (c) Specificity for the first 20 components on a logarithmic vertical axis.}
  \label{fig:ssm_method_comparison}
\end{figure*}

Compactness differed among the four representations (Fig.~\ref{fig:ssm_method_comparison}a). 
Five components retained 89.52\% of coordinate variance for PCA and 13.90\% of total independence measure for ICA. 
Polynomial and RBF KPCA retained 80.67\% and 52.89\% of their respective kernel-feature variance.

Generalization performance was best for PCA among all tested methods at five, ten, and twenty components 
(Fig.~\ref{fig:ssm_method_comparison}b). 
At 20 components, mean RMSE was $2.451 \pm 0.538$~mm for PCA, $3.381 \pm 0.820$~mm for ICA, 
$3.932 \pm 1.538$~mm for polynomial KPCA, and $3.172 \pm 1.035$~mm for RBF KPCA.

Specificity yielded a different ranking from generalization (Fig.~\ref{fig:ssm_method_comparison}c). 
At 20 components, the errors were $10.851 \pm 2.508$~mm for PCA, $6.474 \pm 1.729$~mm for ICA, 
$107.672 \pm 44.021$~mm for polynomial KPCA, and $8.943 \pm 2.009$~mm for RBF KPCA. 
ICA achieved the lowest specificity error, whereas PCA achieved the lowest generalization error. 
The large polynomial-KPCA specificity error indicates unstable preimage extrapolation 
under the chosen sampling range.

\section{Discussions}
To our knowledge, this is the first posture-neutralized parametric model focused on head-neck-shoulder anthropometry that preserves physical size. Existing models mainly describe craniofacial variation in the head or head–neck regions and do not include the shoulders~\cite{zhuang2013head,dai2020statistical}. Our model covers the head, neck, and shoulders in a common estimated cervical pose, providing a foundation for the design of neck-related products.

\subsection{Anthropometry}
Our posture-neutralized model showed greater compactness and lower specificity error than the matched non-neutralized model. Earlier whole-body normalization also increased compactness and reduced body-shape prediction error from anthropometric measurements~\cite{danckaers2019posture}. Posture correction similarly improved the accuracy of a hand SSM relative to a model without correction~\cite{yang2021postureinvariant}. These studies and our results support controlling joint posture before statistical shape analysis. In our model, held-out reconstruction error improved at five components but not at ten. The present framework applies this approach to head-neck-shoulder anthropometry while retaining physical size.

Our PC1 represented coordinated size variation across the head, neck and shoulders, whereas PC2 mainly involved neck circumference and shoulder width. Earlier head models also associated PC1 with overall size~\cite{zhuang2013head,liu2015variations}, but linked PC2 to facial proportions or width~\cite{zhuang2013head,zhang2022age}. Overall size is therefore a shared pattern, while the dominant regional proportions differ. Including the neck and shoulders may therefore shift the leading variation towards these regions.

Our gender-specific regressions linked higher BMI to outward changes around the lower face, neck and shoulders in the predicted mean shapes (Fig.~\ref{fig:sex_bmi_models}). Previous head models showed earlier maturation of head size in girls than boys~\cite{zhang2023modeling} and wider, shorter heads in Chinese than Caucasian populations~\cite{zhang2026growth}. A shared finding is that population characteristics are associated with differences in average geometry. Our results concern BMI-related variation across the adult head-neck-shoulder region, whereas these studies describe head growth and population differences.

\subsection{Applications}
The posture-neutralized HNS shape space is a design input, not a treatment target for an individual user. It can support population coverage analysis, size grouping, representative virtual forms, and parameterized prototypes. These uses extend established applications of 3D anthropometry in sizing systems and product design~\cite{lacko2017product,kuo2020developing}. Designers can select shapes that span the retained variation, inspect the associated surface regions and measurements, and test candidate product geometries before physical prototyping~\cite{luximon2016design}. This approach may reduce reliance on a single mean form or on independent one-dimensional measurements when a product contacts several head, neck, and shoulder regions.

For pillows and cervical supports, the model can provide candidate information on neck height, neck circumference, cervical contour, and the head-neck relationship~\cite{chen2012pillow}. For helmets and other head-worn devices, it can also inform the head dimensions, shoulder clearance, and contact regions that should be considered together. A low-dimensional PCA representation is practical for such iterations because it achieved the lowest held-out reconstruction error among the tested representations. This result supports its use in the present design workflow. It does not establish that PCA covers every valid HNS shape or that a lower specificity error alone would define a better model. The model also does not validate comfort, pressure distribution, wearing stability, or clinical benefit. Those outcomes require product-specific tests~\cite{langley2018comfort,lei2021pillow,pang2018helmet}.

\section{Conclusion}

This study developed a posture-neutralized statistical shape model for the head-neck-shoulder region from three-dimensional scans of 200 Chinese adults. Mapping each scan to a common estimated cervical pose before PCA reduced the mixing of cervical bending with the leading shape modes seen in the matched model without neutralization. Eleven anthropometric measurements made these modes easier to interpret, and PCA had the lowest generalization error among the tested representations. Within this cohort, the neutral-pose shape space may support population coverage analysis, size grouping, representative virtual forms, and parameterized ergonomic prototypes. External cohorts, repeated and multi-posture scans, and product tests are needed to assess transferability, reproducibility, fit, pressure, and comfort.

\section{Acknowledgments}

This work was supported by the National Natural Science Foundation of China (NSFC) Youth Grant (Project No. 52605353), the Innovation and Technology Fund (ITF) through The Hong Kong Research Institute of Textiles and Apparel (HKRITA) (Project No. ITP/016/26TI), the Science and Technology Development Fund of Macau SAR (Project No. 0057/2025/ITP1), Macao Polytechnic University (Project No. RP/FCA-15/2025), and Scientific Research Project of the Health System in Pingshan District, Shenzhen (Project No.2025097).

\bibliography{mybibfile}

\end{document}